\documentclass[a4paper,aps,preprint]{revtex4}

\usepackage{graphicx}
\usepackage{amssymb}

\begin{document}

\title{Sampling of quantum dynamics at long time}

\author{Alessandro Sergi}

\affiliation{
School of Physics, University of KwaZulu-Natal, Pietermaritzburg Campus,
Private Bag X01 Scottsville, 3209 Pietermaritzburg, South Africa }

\author{Francesco Petruccione}

\affiliation{
School of Physics, Quantum Research Group and 
National Institute for Theoretical Physics,
University of KwaZulu-Natal, Westville Campus, Private Bag X54001,
Durban 4000, South Africa }

\begin{abstract}
The principle of energy conservation leads to 
a generalized choice of transition probability
in a piecewise adiabatic representation of quantum(-classical) dynamics.
Significant improvement (almost an order of magnitude, depending on the
parameters of the calculation) over previous schemes is achieved.
Novel perspectives for theoretical calculations in
coherent many-body systems are opened.
\end{abstract}

\maketitle

%%%%%%%%%%%%%%%%%%%%%%%%%%%%%%%%%%%%%%%%%%%%%%%%%%%%%%%%%%%
%Computational techniques
%classical mechanics, 45.10.-b
%continuum mechanics, 46.15.-x
%electronic structure
%atoms and molecules, 31.15.-p
%solids, 71.15.Dx
%fluid dynamics, 47.11.-j
%mathematics, 02.70.-c
%statistical physics and nonlinear dynamics, 05.10.-a
%Quantum jumps, 42.50.Lc
%Quantum mechanics, 03.65.-w
%Quantum statistical mechanics, 05.30.-d

%%%%%%%%%%%%%%%%%%%%%%%%%%%%%%%%%%%%%%%%%%%%%%%%%%%%%%%%%%%%%%%%%%
The importance of theoretical methods for the calculation
of time-dependent quantum properties cannot be emphasized enough.
The lack of general algorithms, so reliable as classical molecular dynamics
simulations~\cite{md}, is to be contrasted with the manifold of open problems
that scientists face both in condensed matter~\cite{photo} and quantum information
technology~\cite{qi}.
Lately, we are also witnessing a renaissance of quantum approaches
to biological phenomena~\cite{physworld}: a revival of interest generated by the combination
of methodologies from open quantum systems~\cite{toqs}
and quantum information theory~\cite{qi}. 
Undoubtedly, the possibility of performing long-time quantum dynamical simulations
would be an asset for all the above fields.

When considering the calculation of time-dependent quantum properties,
two main methods are available: time-dependent density functional theory~\cite{dft}
and quantum-classical formalisms~\cite{book-qc}.
Time-dependent density functional methods are usually limited to linear
response while quantum-classical methods are restricted to perturbations
around the adiabatic evolution, i.e., nonadiabatic corrections,
of few relevant quantum degrees of freedom interacting with a classical bath.
Nevertheless, quantum-classical methods promise to access
the investigation of properties relevant to biological systems\cite{book-qc}.
Here, we are considering the formulation of quantum-classical theory
by means of algebraic brackets which was proposed 
originally in~\cite{qc} and shown to arise from a linear approximation
of the partially Wigner transformed quantum commutator~\cite{kc}.
It is remarkable that, when the environmental degrees of freedom
are harmonic and the coupling to the quantum subsystem is linear in the bath
degrees of freedom, as in gauge theory~\cite{weinberg},
such theory becomes fully quantum.
What is interesting from a computational point of view is that,
within such  a theory, a particular approximation (called
\emph{momentum-jump} in the adiabatic basis of the total system) 
leads to represent  nonadiabatic dynamics
in terms of piecewise (adiabatic) deterministic trajectories
interspersed by stochastic quantum transitions~\cite{sstp,theorchemacc}.

There is a long history of development and methods 
for treating non-adiabatic transitions with 
so called surface-hopping schemes~\cite{book-qc}.
Such schemes were originated in~\cite{sh-orig}. 
A more recent approach can be found in~\cite{heller}.
These methods are successful for the description of the dynamics
but do not easily lead to an accurate formulation
of the statistical mechanics of quantum-classical systems.
Instead, the theory stemming from~\cite{kc} allows one
to address the consistent formulation of the quantum-classical statistical
mechanics~\cite{nielsen} of general hybrid systems, i.e., the theory can describe, 
in the non relativistic limit, \emph{any} quantum subsystem
coupled to a classical bath.
It exactly conserves the energy of the total system
and consistently describes the coupling between the quantum subsystem
and the classical bath (or the quantum harmonic bath represented in Wigner
phase space). The forms of the equations in the momentum-jump approximation
also naturally provide a sampling transition probability for nonadiabatic
change of state. 
However, when nonadiabatic effects are included,
the phase space trajectories representing the quantum(-classical) dynamics
do no longer conserve the energy. Despite this, in its original formulation,
called sequential time-step propagation~\cite{sstp,theorchemacc},
the algorithm is successful, although limited to somewhat short-times
because of numerical instabilities arising from the sampling of the nonadiabatic
transitions. The instability, in practice, restricts the range of applications
of the method to charge transfers and rate processes~\cite{rk-review}.

More general quantum processes require the ability of sampling at
longer time.
In this letter we show how to achieve this by means of a suitable
generalization (implementing the principle of energy conservation)
of the transition probability in the sequential time-step propagation:
this is the main theoretical idea we propose and it is introduced by
Eq.~(\ref{eq:prob-jump}) in the following.
Before providing our solution,
we sketch the theory and the original version of the sequential time-step
propagation, which will be referred to in the following as \emph{primitive} 
algorithm. The interested reader will find more details
in~\cite{theorchemacc}.
The theory of quantum-classical dynamics is defined by
the equation~\cite{b3,b4,ilya}
\begin{eqnarray}
\frac{d}{dt}\hat{\chi}_W(X)
&=&\frac{i}{\hbar}
\left[\begin{array}{cc}\hat{H}_W &\hat{\chi}_W\end{array}\right]
\nonumber\\
&\cdot&
\left[\begin{array}{cc}
0 & 1+\frac{i\hbar}{2}\overleftarrow{\mbox{\boldmath$\partial$}}\cdot
\mbox{\boldmath$\cal B$}^c\cdot \overrightarrow{\mbox{\boldmath$\partial$}}
\\
-1-\frac{i\hbar}{2}\overleftarrow{\mbox{\boldmath$\partial$}}\cdot
\mbox{\boldmath$\cal B$}^c\cdot \overrightarrow{\mbox{\boldmath$\partial$}}
& 0
\end{array}\right]
\nonumber\\
&\cdot&
\left[\begin{array}{c}\hat{H}_W \\ \hat{\chi}_W\end{array}\right]
\;,
\label{eq:whqc-bracket}
\end{eqnarray}
where $\hat{\chi}_W(X)$ is a quantum operator in a partial Wigner
representation depending on the phase space point
$X=(R,P)$, comprised by coordinated and momenta respectively;
$\hat{H}_W$ is the partially Wigner-transformed Hamiltonian operator
of the total system,
$\mbox{\boldmath$\cal B$}^c$ is the symplectic matrix,
and $\mbox{\boldmath$\partial$}$ stands for the phase space gradient
$\partial/\partial X$, with the arrow giving the direction of action
of the operators.
Without loss of generality, one can assume the form
of the Hamiltonian to be 
$
\hat{H}_W(X)=\frac{P^2}{2M}+\hat{h}_W(R)\;.
$
In the adiabatic basis, defined by
the eigenvalue equation
$
\hat{h}_W(R)|\alpha;R\rangle=E_{\alpha}(R)|\alpha;R\rangle\;,
$
the quantum-classical evolution reads
\begin{equation}
\chi_W^{\alpha\alpha'}(X,t)
=\sum_{\beta\beta'}\left(e^{it{\cal L}}\right)_{\alpha\alpha',\beta\beta'}
\chi_W^{\beta\beta'}(X)\;,
\label{eq:qc-evol}
\end{equation}
where
$
i{\cal L}_{\alpha\alpha',\beta\beta'}
=i{\cal L}^0_{\alpha\alpha'}
\delta_{\alpha\beta}\delta_{\alpha'\beta'}
+J_{\alpha\alpha',\beta\beta'}
$.
The diagonal operator $i{\cal L}^0_{\alpha\alpha'}=
\left(i\omega_{\alpha\alpha'}+iL_{\alpha\alpha'}\right)
\delta_{\alpha\beta}\delta_{\alpha'\beta'}$
is defined in terms 
of the quantum adiabatic frequency
$
\omega_{\alpha\alpha'}(R)
=\left(E_{\alpha}(R)-E_{\alpha'}(R)\right)/\hbar
$
and of the classical-like Liouville operator
$
iL_{\alpha\alpha'}=(P/M)\cdot\partial/\partial R
+(1/2)\left(F_W^{\alpha}+F_W^{\alpha'}\right)\cdot
(\partial/\partial P)
$,
where  $F_W^{\alpha}$ are the Hellmann-Feynman forces~\cite{hf-force}.
The operator $J_{\alpha\alpha',\beta\beta'}$ is purely off-diagonal
and its action realizes the quantum nonadiabatic transitions.
It is worth remarking that Eqs.~(\ref{eq:whqc-bracket})
and~(\ref{eq:qc-evol}) exactly conserve the total Hamiltonian
of the system $\hat{H}_W(X)$.

In the momentum-jump approximation, the form of the off-diagonal operator 
$J_{\alpha\alpha',\beta\beta'}$ is simplified~\cite{theorchemacc}.
Here, we denote such an approximation by  $J^{\rm (mp)}_{\alpha\alpha',\beta\beta'}$.
The action of $J^{\rm (mp)}_{\alpha\alpha',\beta\beta'}$ changes the 
quantum state and rescales the bath momenta. The technical details
can be found, among many other possible references, in~\cite{theorchemacc}.
Using the momentum-jump operator, one can also define a momentum-jump
Liouville operator,
$i{\cal L}^{\rm (mp)}_{\alpha\alpha',\beta\beta'}=i{\cal L}^0_{\alpha\alpha',\beta\beta'}
+J^{\rm (mp)}_{\alpha\alpha',\beta\beta'}
$,
approximating the exact operator 
$i{\cal L}_{\alpha\alpha',\beta\beta'}$
in Eq.~(\ref{eq:qc-evol}). 

Deterministic dynamics is too-complicated to be solved, so one has to resort
to stochastic schemes. A very elegant approach is provided by
the sequential short time propagation~\cite{theorchemacc}
(the primitive algorithm).
This is summarized as follows.
For a small time step $\tau$ the quantum-classical propagator
is approximated as
\begin{equation}
\left(e^{i\tau{\cal L}^{\rm (mp)}}\right)_{\alpha\alpha',\beta\beta'}
\approx
e^{i\tau {\cal L}^0_{\alpha\alpha'}}
\left(\delta_{\alpha\beta}\delta_{\alpha'\beta'}
+\tau  J^{\rm (mp)}_{\alpha\alpha',\beta\beta'}\right)\;.
\label{eq:sstp}
\end{equation}
One can prove that the concatenation of short time steps,
according to Eq.~(\ref{eq:sstp}),
reproduces exactly the Dyson integral expansion
of the operator 
$\exp\left(i\tau {\cal L}^{\rm (mp)}\right)_{\alpha\alpha',\beta\beta'}$.
The algorithm unfolds by considering
the action of $J^{\rm (mp)}$ dictated by a stochastic process
with a certain transition probability.
The form of $J^{\rm (mp)}$ naturally suggests
the following \emph{primitive} choice of
the transition probability
(for example considering the $\alpha\to\beta$ quantum transition):
\begin{eqnarray} 
{\cal P}_{\alpha\beta}^{(0)}(X,\tau)
=\frac{|\frac{P}{M}\cdot d_{\alpha\beta}(R)|\tau}
{1+|\frac{P}{M}\cdot d_{\alpha\beta}(R)|\tau}\;,
\label{eq:prob-jump0}
\end{eqnarray} 
where $d_{\alpha\beta}=\langle\alpha;R|\partial/\partial R|\beta;R\rangle$
is the coupling vector.
Normalization fixes the probability of not making any transition
in the time interval $\tau$ as
$
{\cal Q}_{\alpha\beta}^{(0)}(X,\Delta t)=1- {\cal P}_{\alpha\beta}^{(0)}
$.
The stochastic propagation amounts to deterministic trajectory segments,
propagating on single or mean energy surfaces, interspersed by
transitions between energy surfaces.
The transitions break the conservation of energy along
the single trajectory: the conservation is only satisfied in an 
averaged sense in the ensemble.
As one can see from Eq.~(\ref{eq:prob-jump0}), arbitrary shifted momenta
$P'$ can arise from a sampled transition.
As experience has shown, this leads in general to very big denominators 
in the left hand
side of Eq.~(\ref{eq:prob-jump0}). This denominators get multiplied
with each other along the trajectory to give its overall weight
in the stochastic ensemble.
The concatenation of big weights arising from nonadiabatic transitions
produces a numerical instability which has so far limited the application
of the primitive algorithm to somewhat short times.

The principle of energy conservation, which is exactly satisfied
by Eq.~(\ref{eq:whqc-bracket}), guides us in defining a generalized 
transition probability as
\begin{eqnarray} 
{\cal P}_{\alpha\beta}(X,\Delta t)
&=&\frac{\tau|\langle\alpha|\dot{\beta}\rangle|
w\left(c_{\cal E},{\cal E}_{\alpha\alpha',\beta\beta'}\right) }
{1+\tau|\langle\alpha|\dot{\beta}\rangle|
w\left(c_{\cal E},{\cal E}_{\alpha\alpha',\beta\beta'}\right) }
\;,
%%%%%%%%%%%%%%%%%%%%%%%%
\label{eq:prob-jump}
\end{eqnarray} 
where we have defined
$(P/M) \cdot d_{\alpha\beta}(R)=\langle\alpha|\dot{\beta}\rangle$.
Again, normalization provides
$
{\cal Q}_{\alpha\beta}^{(0)}(X,\Delta t)=1- {\cal P}_{\alpha\beta}
$.
Upon introducing the
variation of energy in any quantum transition
$
{\cal E}_{\alpha\alpha',\beta\beta'}
=\frac{P^{\prime 2}}{2M}+
\frac{1}{2}\left(E_{\alpha}(R)+E_{\alpha'}(R)\right)
-\frac{P^2}{2M}
-\frac{1}{2}\left(E_{\beta}(R)+E_{\beta'}(R)\right)
$,
the generalized weight introduced in Eq.~(\ref{eq:prob-jump})
is defined as
\begin{equation}
w\left(c_{\cal E}\right)
=
\left\{
\begin{array}{ccc}
& 1 & {\rm if}~ |{\cal E}_{\alpha\alpha',\beta\beta'}| \le c_{\cal E} ;\\
& 0 & {\rm otherwise};
\end{array}
\right.
\label{eq:eweight}
\end{equation}
with $c_{\cal E}$ tunable constants
controlling the numerical error on the energy conservation.

The generalized transition probability in Eq.~(\ref{eq:prob-jump})
and the energy-conserving weight in Eq.~(\ref{eq:eweight}) are
our fundamental findings, improving the primitive algorithm.
Because of our choice of $w\left(c_{\cal E}\right)$, the sampled
transitions can only allow shifted momenta $P'$ which conserve
(within the required numerical error specified by $c_{\cal E}$)
the energy of the system. This in turn avoids arbitrarily big denominators
in Eq.~(\ref{eq:prob-jump}) and dramatically improves the stability of
the algorithm.

\begin{figure}
%\resizebox{22cm}{16cm}{
\includegraphics* {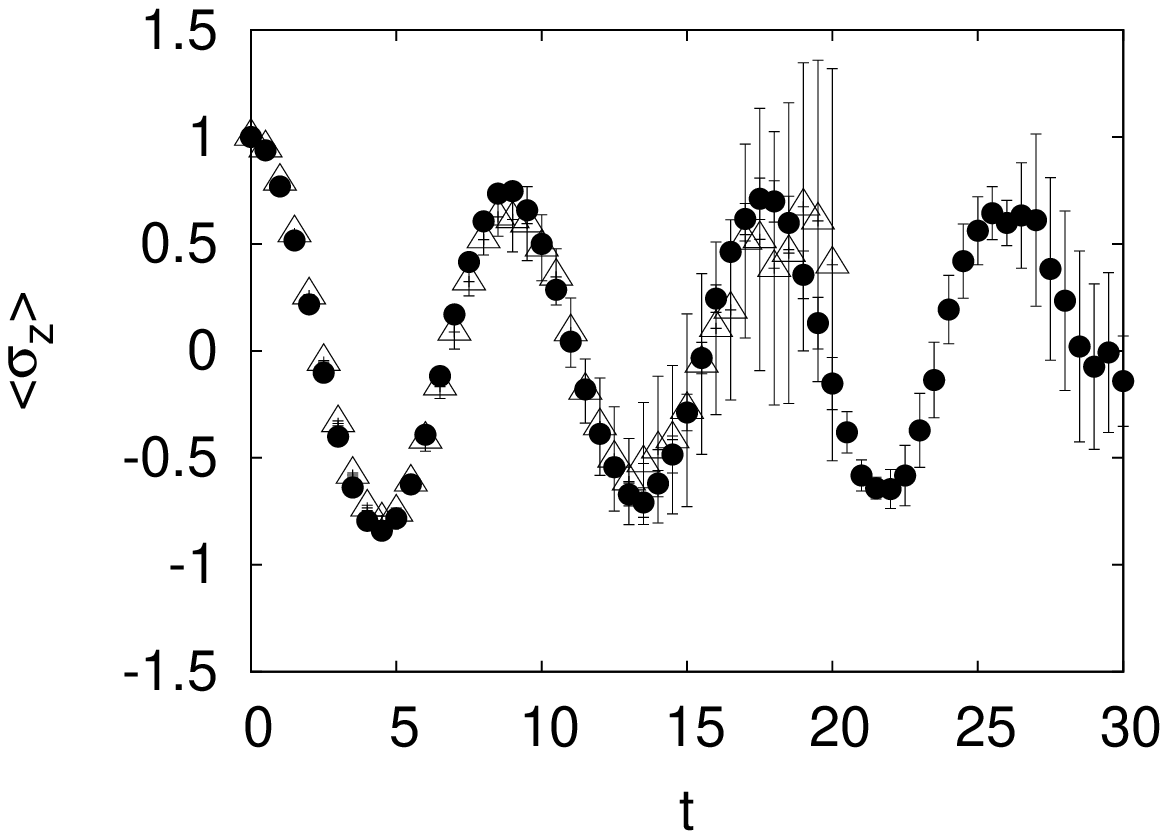}
%}
\caption{
}
Comparison of the primitive ($\triangle$) and energy-conserving
($\bullet$) sampling
for $\beta=0.3$, $\Omega=1/3$, and $\xi=0.007$.
The calculation with the primitive algorithm are propagated
until $t=20$ and then stopped since already around $t=15$
the statistical error becomes very big, as the error bars show.
The calculation with the energy-conserving sampling ($\bullet$),
with $c_{\cal E}=0.01$,
can be extended further than $t=30$.
\label{fig:fig1}
\end{figure}

\begin{figure}
%\resizebox{22cm}{16cm}{
\includegraphics* {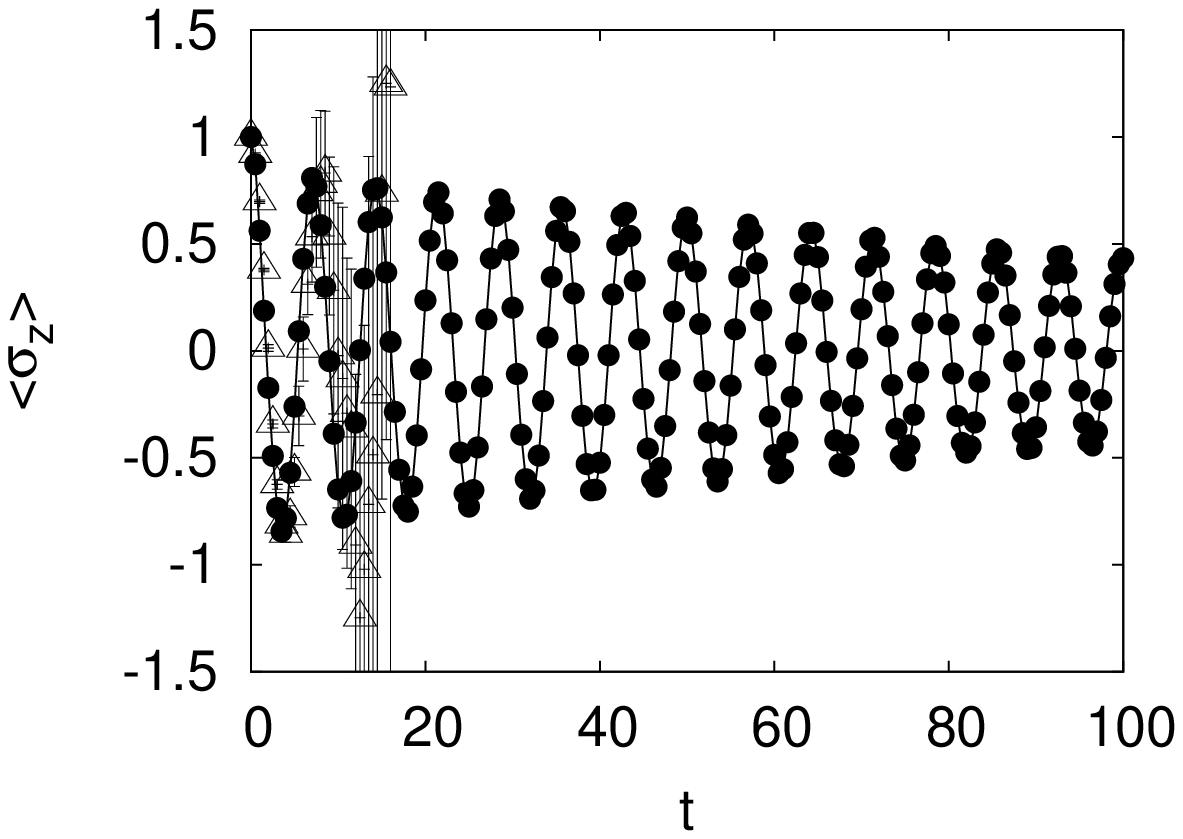}
%}
\caption{
}
Comparison of primitive ($\triangle$) and energy-conserving
($\bullet$) sampling
for $\beta=12.5$, $\Omega=0.4$, and $\xi=0.09$.
The calculation with the primitive algorithm are propagated
until $t=20$ and then stopped since already around $t=12$
the statistical error becomes very big, as the error bars show.
The calculation with the energy-conserving sampling
($\bullet$ and a continuous line to help the eye),
 with $c_{\cal E}=0.01$,
can be extended further than $t=100$ (the error bars of the order of
magnitude of the $\bullet$ symbols).
\label{fig:fig2}
\end{figure}

In order to illustrate the efficiency of the energy-conserving sampling,
we performed a series of calculation on the dynamics of the spin-boson
model~\cite{leggett}. The partially Wigner transformed Hamiltonian of this model
(in scaled coordinates) is $\hat{H}_W=-\Omega\hat{\sigma}_x
+\sum_{j=1}^N(P_j^2/s+\omega_j^2R_j^2/2-c_jR_j\hat{\sigma}_z$,
where $\hat{\sigma}_x$ and $\hat{\sigma}_z$ are the Pauli matrices,
$R_j,P_J$ are coordinates and momenta of $N$ harmonic degrees of freedom
(in the following we have used $N=200$,
$c_j$ are coupling constants defined in term of the Kondo parameter $\xi$).
Details of the model and definition of coordinates and parameters
can be found in~\cite{theorchemacc}.
Figures~\ref{fig:fig1}
 and~\ref{fig:fig2}
illustrate the numerical comparison between the primitive and
our energy-conserving sampling
for the relaxation dynamics
of $\sigma_z$ for various couplings, temperatures, and tunnel splitting $\Omega$.
The results obtained with the primitive algorithm are displayed
by white triangles while those obtained with our energy-conserving sampling
by black filled circles.
Figure~\ref{fig:fig1} displays the results of the numerical calculation
for $\beta=0.3$, $\Omega=1/3$, and $\xi=0.007$. Basically, after $t=15$
(in dimensionless units)
the error bars on the primitive algorithm results grow exponentially fast
and the calculation is stopped at $t=20$. Instead, the calculation with
our generalized sampling scheme can be extended further than $t=30$:
for this set of parameters we obtain an improvement over the time
interval we can sample of at least two.
Figure~\ref{fig:fig2} displays the results of the numerical calculation
for $\beta=12.5$, $\Omega=0.4$, and $\xi=0.09$. This time, the statistical
errors of the primitive algorithm start growing fast around $t=10$,
while our scheme can reach further than $t=100$, providing
an improvement  of an order of magnitude.
Summarizing, the simulation shows that the use of our energy-conserving sampling 
dramatically improves
the stability of the elegant sequential time step algorithm at long time.

In conclusion, it is worth mentioning that the approach, 
embodied by Eq.~(\ref{eq:prob-jump}), to modify
the transition probability in order to respect a conservation law
and improving the stability is very general: it is by no means
restricted to quantum(-classical) dynamics in the 
partial Wigner representation. 
On the contrary, there are reasonable expectations that the generalized scheme
that we have presented here can be applied, after suitable changes,
to other stochastic approaches for calculating time-dependent
properties, both in the classical and quantum cases.

\section*{Acknowledgments}
A. S. is grateful to Professor Raymond Kapral for 
his mentoring in the past years.
This work is based upon research supported by the South
African Research Chair Initiative of the Department of
Science and Technology and National Research Foundation.

%%%%%%%%%%%%%%%%%%%%%%%%%%%%%%%%%%%%

\end{document}